\documentclass{article}

\usepackage{arxiv}
\usepackage{amsmath,amsfonts,amssymb}
\usepackage{algorithmic}
\usepackage{algorithm}

\usepackage[utf8]{inputenc} 
\usepackage[T1]{fontenc}    
\usepackage{hyperref}       
\usepackage{url}            
\usepackage{booktabs}       
\usepackage{nicefrac}       
\usepackage{microtype}      
\usepackage{graphicx}
\usepackage[square,numbers]{natbib}
\usepackage{doi}

\title{Nonconvex optimization for optimum retrieval of the transmission matrix of a multimode fiber}

\date{August 1, 2023}	

\author{ {Shengfu Cheng†}\\
	Department of Biomedical Engineering\\
	The Hong Kong Polytechnic University\\
	Hong Kong SAR, China\\
	\texttt{shengfu.cheng@connect.polyu.hk} \\
    \And
	{Xuyu Zhang†} \\
	Key Laboratory for Quantum Optics\\
	Shanghai Institute of Optics and Fine Mechanics\\
	Chinese Academy of Sciences\\
	Shanghai 201800, China\\
	\texttt{980992193@qq.com} \\
    \AND
	{Tianting Zhong†, Huanhao Li, Haoran Li} \\
	Department of Biomedical Engineering\\
	The Hong Kong Polytechnic University\\
	Hong Kong SAR, China\\
	\texttt{tianting-simon.zhong@connect.polyu.hk}\\
	\texttt{huanhli@polyu.edu.hk}\\
	\texttt{haoran1996.li@connect.polyu.hk} \\
    \And
	{Lei Gong} \\
	Department of Optics and Optical Engineering \\
	University of Science and Technology of China\\
	Hefei 230026, China \\
	\texttt{leigong@ustc.edu.cn} \\
	\And
	{Honglin Liu} \\
	Key Laboratory for Quantum Optics\\
	Shanghai Institute of Optics and Fine Mechanics\\
	Chinese Academy of Sciences\\
	Shanghai 201800, China\\
	\texttt{hlliu4@hotmail.com} \\
	\And
	{Puxiang Lai \thanks{Correspondance should be addressed with hlliu4@hotmail.com and puxiang.lai@polyu.edu.hk} } \\
	Department of Biomedical Engineering\\
	The Hong Kong Polytechnic University\\
	Hong Kong SAR, China\\
	\texttt{puxiang.lai@polyu.edu.hk} \\
}

\renewcommand{\shorttitle}{\textit{arXiv} Template}

\hypersetup{
pdftitle={A template for the arxiv style},
pdfsubject={q-bio.NC, q-bio.QM},
pdfauthor={David S.~Hippocampus, Elias D.~Striatum},
pdfkeywords={First keyword, Second keyword, More},
}

\begin{document}
\maketitle

\begin{abstract}
	Transmission matrix (TM) allows light control through complex media such as multimode fibers (MMFs), gaining great attention in areas like biophotonics over the past decade. The measurement of a complex-valued TM is highly desired as it supports full modulation of the light field, yet demanding as the holographic setup is usually entailed. Efforts have been taken to retrieve a TM directly from intensity measurements with several representative phase retrieval algorithms, which still see limitations like slow or suboptimum recovery, especially under noisy environment. Here, a modified non-convex optimization approach is proposed. Through numerical evaluations, it shows that the nonconvex method offers an optimum efficiency of focusing with less running time or sampling rate. The comparative test under different signal-to-noise levels further indicates its improved robustness for TM retrieval. Experimentally, the optimum retrieval of the TM of a MMF is collectively validated by multiple groups of single-spot and multi-spot focusing demonstrations. Focus scanning on the working plane of the MMF is also conducted where our method achieves 93.6\% efficiency of the gold standard holography method when the sampling rate is 8. Based on the recovered TM, image transmission through the MMF with high fidelity can be realized via another phase retrieval. Thanks to parallel operation and GPU acceleration, the nonconvex approach can retrieve an 8685×1024 TM (sampling rate=8) with 42.3 s on a regular computer. In brief, the proposed method provides optimum efficiency and fast implementation for TM retrieval, which will facilitate wide applications in deep-tissue optical imaging, manipulation and treatment.
\end{abstract}

\keywords{Transmission matrix \and Phase retrieval \and Multimode fiber imaging \and Wavefront shaping}

\section{Introduction}
Different from ordinary ballistic optics, light propagation in complex media is highly disordered \citenum{ref1, ref2} due to the multiple scattering occurring in media like biological tissues or mode dispersion in multimode fiber (MMF). Finding an order out of such disorders has been long pursued. Over the past decade, enormous progresses have been made via wavefront shaping \citenum{ref3, ref4, ref5, ref6, ref7, ref8} and especially the transmission matrix (TM) method \citenum{ref9, ref10, ref11, ref12, ref13} in controlling light to focus and image through complex media. The TM of a disordered medium describes the complex output responses for an arbitrary point-source input, which is regarded as the transfer function of the medium under the shift-invariance assumption \citenum{ref11}. The measurement of TM offers a versatile tool to control light delivery in spite of scattering \citenum{ref6, ref10}, as well as recovering object information from acquired speckle patterns \citenum{ref14, ref15}. The TM method has spurred a wide range of MMF-based applications, such as focusing \citenum{ref16, ref17}, glare suppression \citenum{ref18, ref19}, endoscopic imaging \citenum{ref20, ref21, ref22}, manipulation \citenum{ref23}, optogenetics \citenum{ref24}, and communication \citenum{ref14, ref25, ref26}.

TM measurement of a scattering medium was first introduced by Popoff \textit{et.al.} \citenum{ref9, ref10} using coaxial holography with internal reference. Since then, various forms of TM measurement have been developed. Typically, the TM is measured by recording the complex output fields under a sequence of input modulations. The modulation basis is usually orthogonal, which can be of diverse forms, including Hadamard matrix \citenum{ref9, ref10}, Fourier transform matrix \citenum{ref27}, point source \citenum{ref28, ref29}, and random phase \citenum{ref13}. Regardless of the form, the measured TM relates all input modes to each output mode by linear superposition. Depending on the type of input modulation and output measurement, the TM could be complex-valued \citenum{ref9, ref29}, real-valued \citenum{ref30, ref31} or even binary \citenum{ref32}. Among them, complex TM is used most as it supports both amplitude and phase modulation of light, which, however, usually entails holographic setup. Off-axis holography based on either phase-shifting \citenum{ref33} or spatial filtering \citenum{ref34} can acquire the complex TM accurately. Nevertheless, effective off-axis interferometry with an additional reference beam and high system stability it demands could be unpracticable in some scenarios. As an example, the coherence length of pulse laser could be too short to be used for interferometry-based TM measurement. With coaxial holography, the above issues might be alleviated, but the dark spot problem with the measured TM caused by speckle reference field \citenum{ref35} is still unsatisfying.

Recent efforts have sought to accurately retrieve the complex TM from intensity-only measurements by using advanced phase retrieval algorithms \citenum{ref13, ref27, ref36, ref37, ref38, ref39, ref40, ref41, ref42}, which started with a Bayesian inference approach (\textit{i.e.}, prVBEM) \citenum{ref13}. This was followed by prSAMP \citenum{ref36} and prVAMP \citenum{ref37}. Although robust to noise, a prior knowledge of noise statistics is a must for these approaches. Semidefinite programming (SDP) that uses convex relation has also been introduced for solving the TM retrieval problem \citenum{ref38}, but it usually requires $N\ln N$  ($N$ is the input size) measurements and tends to be computationally inefficient. Additionally, works based on extended Kalman filter (EKM) \citenum{ref39} or generalized Gerchberg-Saxton (GGS) algorithm \citenum{ref40} claim retrieving TM with  measurements could be enough. That said, EKM is computationally burdened and hard for parallelization. GGS is efficient in computation, but its performance is still suboptimum in real practice. Most recently, the area also sees the birth of a smoothed Gerchberg-Saxton algorithm \citenum{ref42} and a nonlinear optimization method \citenum{ref27} for TM retrieval.

To overcome the aforementioned limitations, in this study a state-of-the-art nonconvex optimization approach is adopted and modified for TM retrieval with optimum performance. Compared with existing TM retrieval algorithms, the proposed nonconvex method can achieve optimal efficiency numerically with less running time or sampling rate. In the experiment, by focus-scanning across the field-of-view (FOV) of an MMF with the acquired TM, the performance of the proposed method is validated to approach the golden standard, i.e., off-axis holography with a sampling rate of 8. Moreover, with the assistance of parallel operation and GPU acceleration, multiple rows of TM can be recovered rapidly. Our method for TM retrieval is featured with optimum efficiency and fast implementation in a reference-less and robust setting, which holds potential for many deep-tissue imaging and focusing applications with the usage of MMF.

\section{Methods}
\subsection{Formulation of the TM retrieval problem}

The theoretical model of retrieving a TM under a sequence of input modulations is formulated as follows. Suppose the number of discrete modulation units (\textit{i.e.}, input size) and speckle field pixels (\textit{i.e.}, output size) is $N$ and $M$ , respectively. Given $P$ calibration patterns such that the probe matrix ${\boldsymbol{X}} \in {\mathbb{C}^{N \times P}}$ and the amplitude measurements ${\boldsymbol{Y}} \in {\mathbb{R}_{+}^{M \times P}}$, the TM ${\boldsymbol{A}} \in {\mathbb{C}^{M \times N}}$ that needs to be estimated shall follow
\begin{equation}
  {\boldsymbol{Y}} = \left| {{\boldsymbol{AX}}} \right|,
  \label{eq:one}
\end{equation}
where takes the absolute value for the elements inside. By taking the conjugate transpose of both sides of Eq.~(\ref{eq:one}), we have
\begin{equation}
  {{\boldsymbol{Y}}^H} = \left| {{{\boldsymbol{X}}^H}{{\boldsymbol{A}}^H}} \right|,
  \label{eq:two}
\end{equation}
where $\left( {  \cdot  } \right)^H$ is the element-wise conjugate transpose operator. Column-wisely, ${\boldsymbol{Y}}^H = \left[ {{{\boldsymbol{y}}_1},{{\boldsymbol{y}}_2}, \cdots ,{{\boldsymbol{y}}_M}} \right]$, where ${\boldsymbol{y}_i} \in {\mathbb{R}_{+}^{P}}$ that constitutes the measurements associated with the ${i^{th}}$ output mode; ${\boldsymbol{A}}^H = \left[ {{{\boldsymbol{a}}_1},{{\boldsymbol{a}}_2}, \cdots ,{{\boldsymbol{a}}_M}} \right]$ where ${{\boldsymbol{a}}_i} \in {\mathbb{C}^N}$ that denotes the ${i^{th}}$  row of TM (after conjugate transpose), $i = 1, \cdots ,M$. In this case, the TM retrieval problem is decomposed into $M$ independent sub-problems given by
\begin{equation}
  {{\boldsymbol{y}}_i} = \left| {{{\boldsymbol{X}}^H}{{\boldsymbol{a}}_i}} \right|, i = 1, \cdots ,M
  \label{eq:three}
\end{equation}
Due to the operation of taking absolute values in Eq.~(\ref{eq:three}), the above problem of estimating one row of TM is nonlinear and falls in the category of phase retrieval. 

Phase retrieval problem has been studied intensively in mathematics as it is commonly encountered in practice, with representative algorithms including alternating projection \citenum{ref43} (\textit{e.g.}, Gerchberg-Saxton and Fineup), SDP \citenum{ref44} (\textit{e.g.}, PhaseLift, PhaseMax, PhaseCut), approximate message passing (\textit{e.g.}, GAMP \citenum{ref45}, VAMP \citenum{ref46}), and nonconvex optimization \citenum{ref47, ref48, ref49, ref50, ref51} etc. Among these methods, nonconvex approaches are proven to be superior and have been developed rapidly in the past years. There are mainly two categories of nonconvex approaches, the intensity-flow \citenum{ref48}, \citenum{ref52} and amplitude-flow models \citenum{ref49, ref50, ref51}, with the latter being better in both empirical success rate and convergence property. In particular, the amplitude-flow models have been proven to converge linearly to the true solution under $O(n)$ Gaussian measurements for a signal with dimension \citenum{ref51}.

\subsection{The modified RAF algorithm}
\label{sect:title}
Herein, the cutting-edge reweighted amplitude flow (RAF) algorithm \citenum{ref51} is adopted for the TM retrieval problem. Solving Eq. (3) can be recast as an optimization problem 
\begin{equation}
  \mathop {{\rm{min }}}\limits_{{{\boldsymbol{a}}_i} \in {\mathbb{C}^N}} L\left( {{{\boldsymbol{a}}_i}} \right) = \left\| {\left| {{{\boldsymbol{X}}^H}{{\boldsymbol{a}}_i}} \right| - {{\boldsymbol{y}}_i}} \right\|_2^2,
  \label{eq:four}
\end{equation}
where $L\left( {{{\boldsymbol{a}}_i}} \right)$ is an amplitude-based least square error (LSE) loss function. While most nonconvex algorithms contain two stages, \textit{i.e.}, spectral initialization and gradient descent, RAF applies reweighting techniques in both stages that accelerates the signal recovery significantly. Considering Eq.~(\ref{eq:four}), the signal, \textit{i.e.}, one row of TM ${\boldsymbol{a}}$ (the row index $i$ is omitted for brevity, same below) is first estimated with the weighted maximum correlation initialization. A subset of the row vectors in the probe matrix (\textit{i.e.}, $S \subset \left\{ {1, \cdots ,P} \right\}$ of the row vectors in the probe matrix (${{\boldsymbol{X}}^H} = \left[ {{\boldsymbol{x}}_1^H;{\boldsymbol{x}}_2^H; \cdots {\boldsymbol{x}}_p^H} \right]$) that correspond to the $\left| S \right|$ largest entries among the measurements ${\boldsymbol{y}} = {\left\{ {{y_j}} \right\}_{1 \le j \le P}}$ are selected. These are called direction vectors, as they are more correlated to the true signal. The signal estimate can be constructed by maximizing its correlation to the direction vectors $\left\{ {{\boldsymbol{x}}_j^H\left| {j \in S} \right.} \right\}$ such that
\begin{equation}
  \mathop {\max }\limits_{\left\| {\boldsymbol{a}} \right\| = 1} \frac{1}{{\left| S \right|}}{\sum\limits_{j \in S} {\left| {\left\langle {{\boldsymbol{x}}_j^H,{\boldsymbol{a}}} \right\rangle } \right|} ^2} = \mathop {\max }\limits_{\left\| {\boldsymbol{a}} \right\| = 1} {{\boldsymbol{a}}^H}\left( {\frac{1}{{\left| S \right|}}\sum\limits_{j \in S} {{{\boldsymbol{x}}_j}{\boldsymbol{x}}_j^H} } \right){\boldsymbol{a}}.
  \label{eq:five}
\end{equation}
By weighting more to the selected ${\boldsymbol{x}}_j^H$ vectors corresponding to larger ${y_j}$ values with the weights $y_j^\alpha ,\forall j \in S$  (exponent $\alpha=0.5$, by default), the solution ${{\boldsymbol{\tilde a}}^0}$ of Eq.~(\ref{eq:five}) is the unit-norm principle eigenvector of the Hermitian matrix: 
\begin{equation}
  {\boldsymbol{H}} = \frac{1}{{\left| S \right|}}\sum\limits_{j \in S} {y_j^\alpha {{\boldsymbol{x}}_j}{\boldsymbol{x}}_j^H}  = {\boldsymbol{X}} \cdot diag(\tilde y_1^\alpha ,\tilde y_2^\alpha , \cdots \tilde y_p^\alpha ) \cdot {{\boldsymbol{X}}^H},
  \label{eq:six}
\end{equation}
where  $\tilde y_j^\alpha  = \left\{ {\begin{array}{*{20}{c}}
  {y_j^\alpha {\rm{,    }}j \in S}\\
  {0{\rm{, otherwise}}}
  \end{array}} \right.$. ${{\boldsymbol{\tilde a}}^0}$ is then scaled to obtain the signal initial guess ${{\boldsymbol{a}}^0} = \sqrt {\sum\nolimits_{j = 1}^P {y_j^2/P} }  \cdot {{\boldsymbol{\tilde a}}^0}$. 
The initialized signal ${\boldsymbol{a}}^0$ is further refined by reweighted “gradient-like” iterations. The gradient of the LSE loss in Eq.~(\ref{eq:four}) with respect to  $\boldsymbol{a}$ is
\begin{equation}
  \nabla L({\boldsymbol{a}}) = \frac{1}{P} \cdot {\boldsymbol{X}}\left( {{{\boldsymbol{X}}^H}{\boldsymbol{a}} - {\boldsymbol{y}} \circ \frac{{{{\boldsymbol{X}}^H}{\boldsymbol{a}}}}{{\left| {{{\boldsymbol{X}}^H}{\boldsymbol{a}}} \right|}}} \right),
  \label{eq:seven}
\end{equation}
where $ \circ $ denotes element-wise multiplication. Let $t$ be the iteration index, then the gradient descent is described by
\begin{equation}
  {{\boldsymbol{a}}^{t + 1}} = {{\boldsymbol{a}}^t} - \mu  \cdot \nabla L({{\boldsymbol{a}}^t}),
  \label{eq:eight}
\end{equation}
where $\mu$ is the step size. One can reweight the gradients in Eq.~(\ref{eq:seven}) that have larger $\left| {{{\boldsymbol{X}}^H}{{\boldsymbol{a}}^t}} \right|\oslash {\boldsymbol{y}}$ ($\oslash$ represents element-wise division), which are deemed more reliable in directing to the true signal. The adaptive weight vector is: 
\begin{equation}
  {\boldsymbol{w}} = \left| {{{\boldsymbol{X}}^H}{{\boldsymbol{a}}^t}} \right|\oslash \left( {\left| {{{\boldsymbol{X}}^H}{{\boldsymbol{a}}^t}} \right| + \beta {\boldsymbol{y}}} \right),
  \label{eq:nine}
\end{equation}
where $\beta$ is a pre-selected parameter. The above descriptions show the reweighted gradient flow for TM retrieval. 

Inspired by Ref. \citenum{ref40}, herein we modify the gradient-descent process of the original RAF algorithm, which is divided into two steps. In Step 1, the normalized measurement vector ${\boldsymbol{y}}$ is replaced with its square ${{\boldsymbol{y}}^2}$ for gradient computation, which enlarges the gradient and functions as the artificial heat data. In Step 2, still the amplitude measurement ${\boldsymbol{y}}$ is used. Step 1 contains the first $2/3$ total iterations, which is set empirically (the rationale is referred to Appendix A). The modified algorithm is simple yet surprisingly effective, which is named RAF 2-1 and shown to reduce the measurement error to a much lower level than the original RAF (see Appendix B). The pseudocode of retrieving one row of TM  with RAF 2-1 is summarized in Algorithm~\ref{alg1}. Note that multiple rows of TM can easily be retrieved in a parallel way. 

\begin{algorithm}
    \renewcommand{\algorithmicrequire}{\textbf{Input:}}
	  \renewcommand{\algorithmicensure}{\textbf{Output:}}
    \caption{RAF 2-1 for retrieving one row of TM, $\boldsymbol{a}$}
    \label{alg1}
    \begin{algorithmic}[1] 
      \STATE  \textbf{Input}: Data $\boldsymbol{y} \in \mathbb{R}_{+}^{P}$ with $\left\{ y_j\right\}_{1 \le j \le P}, \boldsymbol{X} \in \mathbb{C}^{N \times P};$ number of iterations $T$; step size $\mu = 3$ and weighting parameter $\beta = 5$; subset cardinality $\left | S \right | = \left \lfloor  3P/13 \right \rfloor$, and exponent $\alpha = 0.5$.   
      
      \STATE  \textbf{Construct} $S$ to include indices associated with the $\left | S \right |$ largest entries among $\left\{ y_j\right\}_{1 \le j \le P}$.
          
      \STATE  \textbf{Initialization}: Let  $\boldsymbol{a}_0:=\sqrt{\sum_j^P{y_j^2}/P} \cdot \tilde{\boldsymbol{a}}_0$  where $\tilde{\boldsymbol{a}}_0$ is the unit-norm principle eigenvector of the Hermitian matrix 
          $$
          \boldsymbol{H} := \boldsymbol{X} \cdot diag(\tilde{y}_1^\alpha, \tilde{y}_2^\alpha, \cdots, \tilde{y}_P^\alpha) \cdot  \boldsymbol{X}^H,
          $$ 
          where $\tilde{y}_j^\alpha := \left\{\begin{matrix}
          y_j^\alpha, j\in S\\ 
          0, otherwise
          \end{matrix}\right.$.
          
      \STATE  \textbf{Gradient-descent loop} \\
          Step 1: for $t=0$ to $\left \lfloor \frac{2}{3}T\right \rfloor-1$ do
          $$
          \boldsymbol{a}^{t+1} = \boldsymbol{a}^{t} - \frac{\mu}{P} \cdot \boldsymbol{X} \left [ \boldsymbol{w} \circ (\boldsymbol{X}^H\boldsymbol{a}^t - \boldsymbol{y^2} \circ  \frac{\boldsymbol{X}^H\boldsymbol{a}^t}{\left | \boldsymbol{X}^H\boldsymbol{a}^t \right |})\right ]
          $$
          Step 2: for $t=\left \lfloor \frac{2}{3}T\right \rfloor$ to $T-1$ do
          $$
          \boldsymbol{a}^{t+1} = \boldsymbol{a}^{t} - \frac{\mu}{P} \cdot \boldsymbol{X} \left [ \boldsymbol{w} \circ (\boldsymbol{X}^H\boldsymbol{a}^t - \boldsymbol{y} \circ  \frac{\boldsymbol{X}^H\boldsymbol{a}^t}{\left | \boldsymbol{X}^H\boldsymbol{a}^t \right |})\right ]
          $$
          where $\boldsymbol{w} := \left |\boldsymbol{X}^H\boldsymbol{a}^t \right | \oslash (\left |\boldsymbol{X}^H\boldsymbol{a}^t \right | + \beta\boldsymbol{y})$.
      \STATE  \textbf{Output}: $\boldsymbol{a}$.
    \end{algorithmic}
\end{algorithm}

\section{Results}
\subsection{Numerical evaluation}

Numerical evaluations were conducted at first to assess the efficiency of the proposed RAF 2-1, with comparisons with several representative TM retrieval algorithms including the pioneering prVBEM and the recent GGS 2-1 that outperformed previous ones. Note that unless otherwise specified, all the following simulations were conducted using a PC with an Intel Xeon CPU (3.50 GHz, 6 cores) and 64 GB RAM in the environment of MATLAB 2022a. For each algorithm, the performance is evaluated by the efficiency of focusing with the retrieved TM, which is indicated by the peak-to-background ratio (PBR). This is performed by taking the conjugate of one row of TM to construct the phase mask for focusing light onto the corresponding position. The TM was modelled using the speckle field produced by random phase mask with zero-padding in the Fourier domain, whose elements obeyed a circular Gaussian distribution. According to Ref. \citenum{ref3}, the theoretical PBR of focusing is linearly proportional to the input size $N$, as given by
\begin{equation}
  \eta  = \frac{\pi }{4}(N - 1) + 1.
  \label{eq:ten}
\end{equation}
The focusing efficiency a certain TM retrieval algorithm can achieve is typically determined by the iterations (or the running time) it has taken and the sampling rate ($\gamma  = P/N$).

\begin{figure}
	\centering
	\includegraphics[width=15cm]{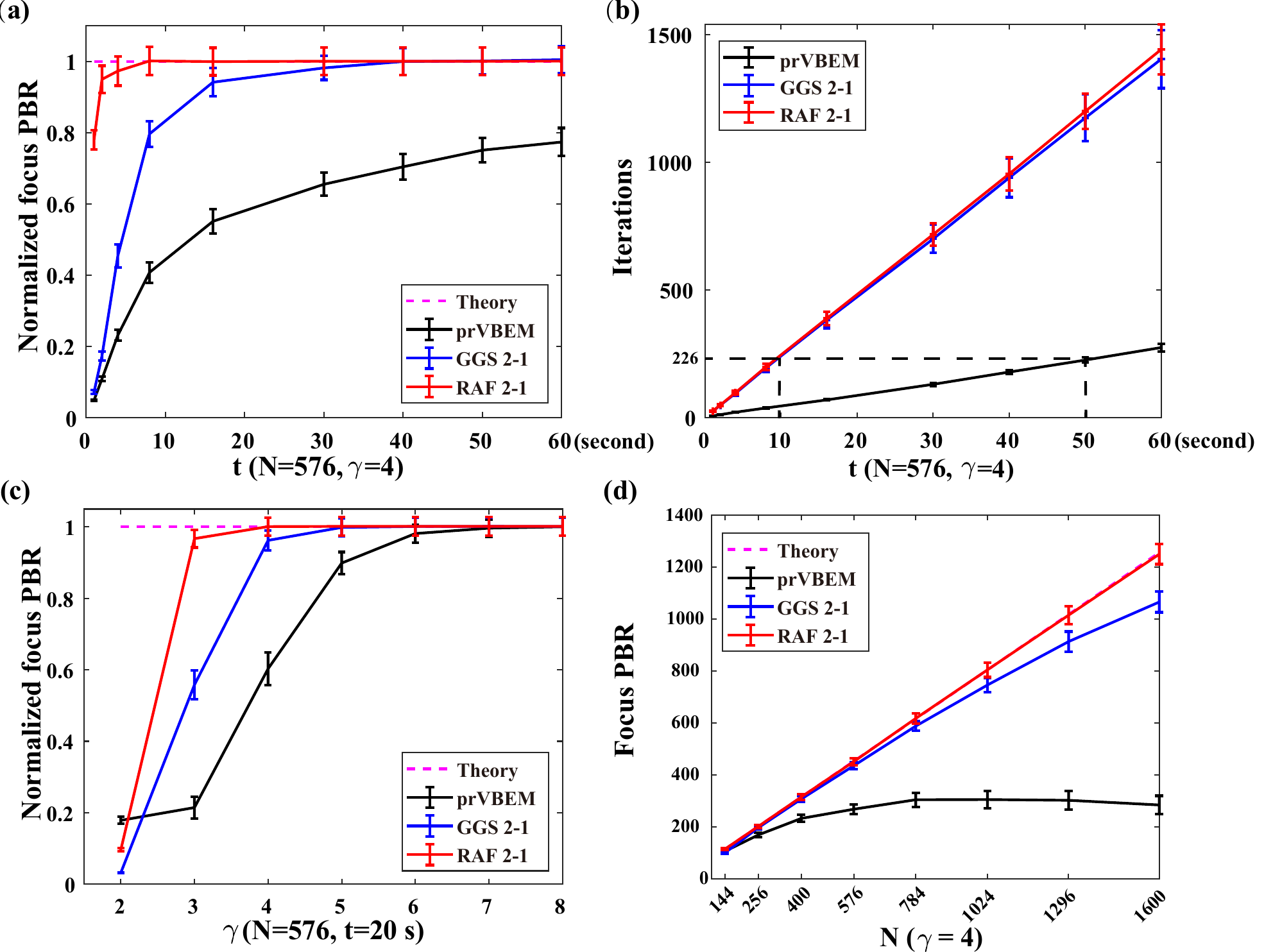}
	\caption{Simulation comparison of TM retrieval performances among prVBEM, GGS 2-1, and RAF 2-1 using the average focus PBRs at positions in the output field with the retrieved TMs. (a) Normalized PBR achieved by different algorithms under a range of running times when $N=576$ and $\gamma=4$ . (b) The iterations taken by different algorithms versus running times for the case of (a). (c) Normalized PBR achieved by different algorithms under a range of $\gamma$ when $N=576$ and a running time of 20 seconds. (d) Focus PBR achieved by different algorithms (with the same running time) under a range of $N$ when $\gamma=4$. Note the error bars denote the standard deviations of 30 repeated tests.}
	\label{fig1}
\end{figure}

\begin{figure}
  \centering
  \includegraphics[width=14cm]{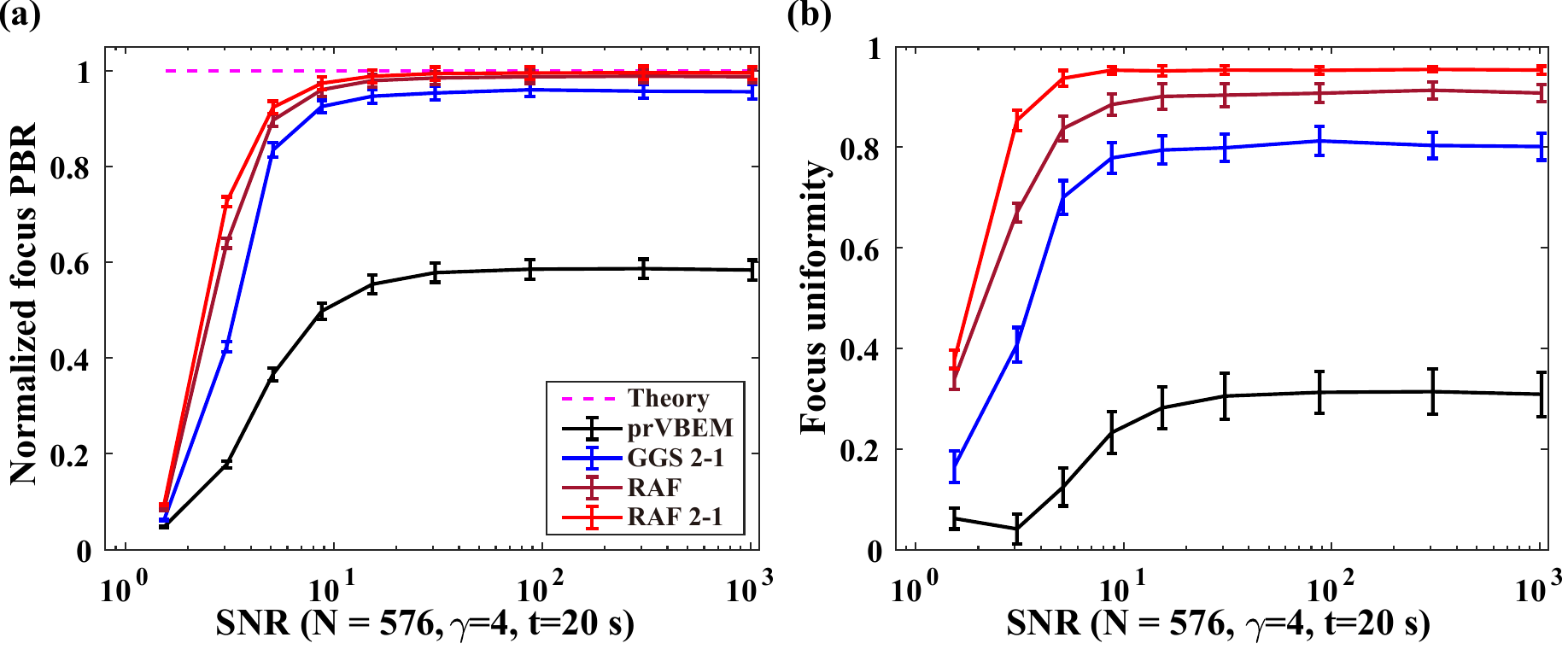}
  
  \caption 
  { \label{fig2}
  Theoretical comparison of TM retrieval performance under measurement noise among different algorithms. (a) Normalized focus PBR and (b) focus uniformity at $20 \times 20$ positions in the output field under various SNR levels when $N=576$, $\gamma=4$, and a running time of 20 s, using the TMs retrieved by prVBEM, GGS 2-1, RAF, and RAF 2-1, respectively. Note the error bars denote the standard deviations of 30 repeated tests.} 
\end{figure}

We first examined the focusing performance of different TM retrieval algorithms under a range of running times when the sampling rate was fixed to be $\gamma=4$. The input phase mask had the size of $24 \times 24$. A total of $20 \times 20$ foci were produced sequentially that corresponded to 400 rows of TM to be retrieved, with the average PBRs of the foci as the focusing PBR. Fig.~\ref{fig1}a presents the average focusing results achieved with a running time ranging from 1 to 60 s based on 30 repeated tests. It can be seen RAF 2-1 reaches the optimum efficiency after running for $\sim8$ s, while GGS 2-1 requires much longer running time ($\sim40$ s) and prVBEM cannot fully approach the optimum within 60 s. This indicates our nonconvex method is superior in algorithm convergence given the same condition. Fig.~\ref{fig1}b additionally shows the number of iterations versus running times, which reveals the seconds per iteration for prVBEM, GGS 2-1, and RAF 2-1 are roughly 5:1:1 in such a case. Hence, the nonconvex approach is at least as highly efficient as GGS 2-1 in computation time, and both are much better than prVBEM. 

The influence of sampling rate was also explored for TM retrieval algorithms, by fixing the running time to be 20 s when $N=576$. As seen in Fig 1c, all algorithms can achieve a higher efficiency of focusing with a larger $\gamma$, as more measurements allow more accurate phase retrieval. Moreover, our RAF 2-1 outperforms its competing peers as it averagely realizes more than 95\% efficiency when $\gamma=3$ and 100 \% when $\gamma=4$. By comparison, GGS 2-1 requires a sampling rate of 5, while prVBEM requires 7 for the optimal performance under the same condition. 

After the study on the effects of running time and sampling rate, the focusing performances of different algorithms under a series of input sizes were further investigated. For fair comparison, the sampling rate was fixed at $\gamma=4$. The running time for all algorithms was the same at each case of $N$, while also increased accordingly with a larger $N=576$ from 144 to 1600, so that the number of iterations taken by prVBEM remained the same as the case of $N=576$ with a running time of 20 s. We can observe from Fig.~\ref{fig1}d that RAF 2-1 always approach the optimum at any input size, while GGS 2-1 and especially prVBEM gradually see a larger and larger distance from the optimum with increased $N$. This is reasonable as the accurate recovery of TM is more demanding for the running time and particularly the sampling rate when at a larger dimension. Since $\gamma=4$ is fixed, prVBEM may step into a plateau and decays slightly in the focusing efficiency starting from $N=784$, as it could not handle a larger signal recovery problem with limited (and insufficient) measurements.

Since measurements are inevitably contaminated with noise in practice, a good noise-robustness is preferred for a TM retrieval algorithm. Thus, the algorithm performances were also evaluated under a range of signal-to-noise (SNR) levels using simulated noisy measurements. In the simulation, a multiplicative noise is added to the output field intensity ${\boldsymbol{I}} \in {\mathbb{R} _ + ^P}$ . The SNR is defined as
\begin{equation}
{\rm{SNR}} = \left\langle {\boldsymbol{I}} \right\rangle /\sqrt {\left\langle {{{\left( {{\boldsymbol{\varepsilon }} - \left\langle {\boldsymbol{\varepsilon }} \right\rangle } \right)}^2}} \right\rangle },
\label{eq:eleven}
\end{equation}
where ${\boldsymbol{\varepsilon }} = {{\boldsymbol{I}}_{{\rm{noise}}}} - {\boldsymbol{I}}$, which denotes the noise vector, ${\boldsymbol{I}}_{{\rm{noise}}}$ is the noisy measurement vector, and $\left\langle  \cdot  \right\rangle $ takes the average for the elements inside. For the focusing results of multiple foci, uniformity is an important metric to measure the focus quality. The focusing uniformity ($\mu$) is given by
\begin{equation}
u = 1 - \sqrt {\left\langle {{{\left( {{I_k} - \left\langle {{I_k}} \right\rangle } \right)}^2}} \right\rangle } /\left\langle {{I_k}} \right\rangle , k \in K
\label{eq:twelve}
\end{equation}
where $K$ is a set of the indexes of foci. With parameter settings that $N=576$, $\gamma=4$ and a running time of 20 s, the results of the normalized PBR and the uniformity of 400 foci produced using different TM retrieval algorithms are given in Fig 2(a-b). Note the original RAF was also included to highlight the improved anti-noise capability by our modification. It is found that a maximum improvement of $\sim9$\% in terms of the focusing efficiency can be realized by RAF 2-1 over RAF under noise. Such a difference between RAF 2-1 and RAF weakens when SNR increases, and their focusing efficiencies are almost the same when SNR is about 30. This explains why RAF was excluded in the previous noiseless tests. Besides, an obvious improvement of the focus uniformity is achieved by RAF 2-1, which is at best $\sim18$\% higher than RAF under noisy conditions, and $\sim5$\% better when SNR is sufficiently large. Overall, algorithm performances in both focusing PBR and uniformity follows the order: RAF 2-1 > RAF > GGS 2-1 > prVBEM. The difference between GGS 2-1 and RAF is relatively small, whereas prVBEM falls behind GGS 2-1 considerably.

\subsection{Experiment}
\begin{figure}
  \centering
  \includegraphics[width=12cm]{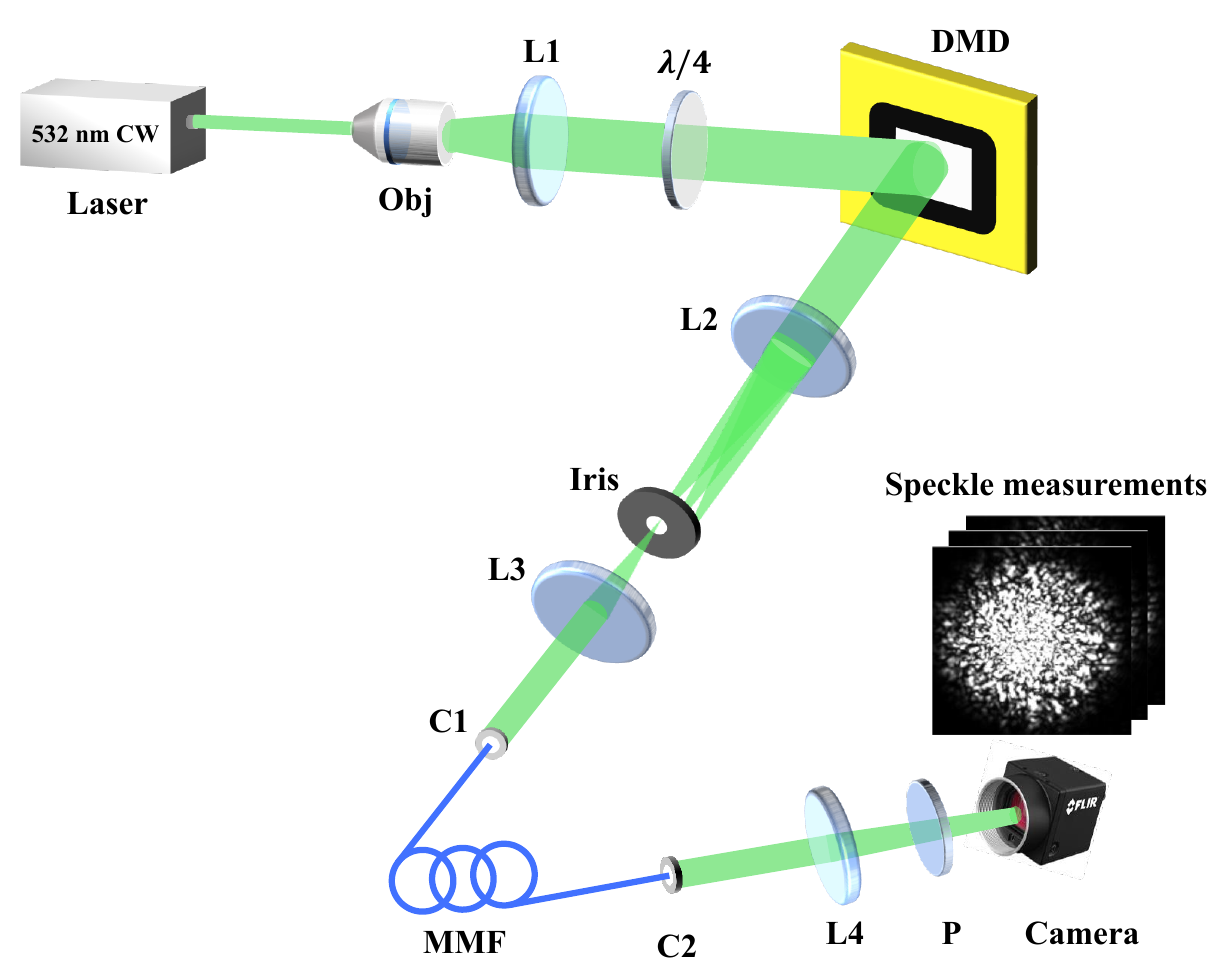}
  
  \caption 
  { \label{fig3}
  Experimental configuration for retrieving the TM of a MMF with speckle intensity measurements. C, collimator; DMD, digital mirror device; L, lens; P, polarizer; MMF, multimode fiber; Obj, objective lens. $\lambda/4$, quarter-wave plate.} 
\end{figure}

The experimental configuration for retrieving the TM of an MMF is illustrated in Fig.~\ref{fig3}. A beam from a 532 nm continuous-wave laser (EXLSR-532-300-CDRH, Spectra Physics, USA) was expanded by a 40x objective lens and collimated by a lens (L1). The linearly polarized beam was then modulated into circularly polarized by a quarter-wave plate ($\lambda/4$) before impinging onto a digital micromirror device (DMD, DLP9500, Texas Instruments Inc, USA). Based on the Lee hologram technique, the DMD, combined with a 4f system composed of L2, iris, and L3, allowed for phase modulation at a high-speed pattern rate of up to 23 KHz, rendering fast data acquisition for TM calibration. The phase-encoded and shrunk beam was subsequently coupled into a MMF of 0.22 numerical aperture (NA) and diameter of 105 $\mu m$ (SUH105, Xinrui, China) by a collimator (C1). The transmitted light was imaged with a collimator (C2), too. The beam then passed through a lens (L4) for convergence and a polarizer before being captured by a CMOS camera (BFS-U3-04S2M, FLIR, USA). For TM retrieval, there was a sequence of speckle intensity measurements under the input modulations of random phase.

\begin{figure}
  \centering
  \includegraphics[width=15cm]{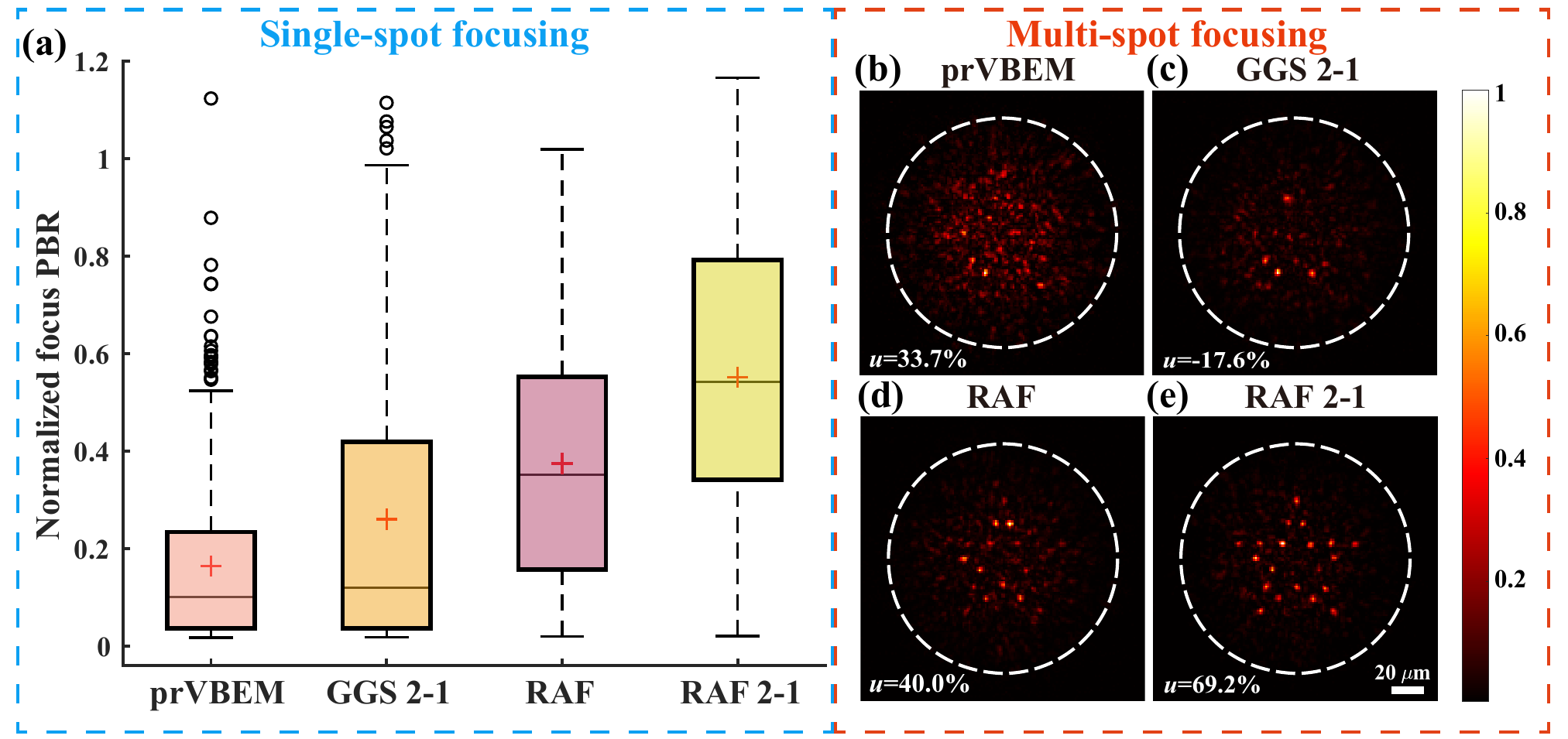}
  
  \caption 
  { \label{fig4}
  Comparison of light focusing results through MMF with the TMs retrieved by different algorithms. (a) The histograms of normalized PBR of $20 \times 20$ foci and (b) the results of multi-spot focusing (pentagram) in the output field of MMF, both obtained by prVBEM, GGS 2-1, RAF, and RAF 2-1 respectively with $N=1024$ and $\gamma=5$. Note the crosses in (a) represent the mean values, the white dashed circles in (b) show the fiber region. The scale bar in (b-e) was 20 $\mu m$.} 
\end{figure}

In experiment, the performances of using different TM retrieval algorithms to control light delivery despite scattering were compared from the aspects of single-spot and multi-spot focusing through MMF. For single-spot focusing, a total of $20 \times 20$ foci were generated sequentially on the working plane of the MMF, which meant 400 rows of TM were to be retrieved. The sampling rate was set to be 5 for all algorithms to ensure the quality of TM retrieval, given that the acquired speckle data suffered from noises such as shot noise, dark current noise, and readout-noise. Fig.~\ref{fig4}a presents the histogram distribution of focusing PBR with different algorithms. Since the experimentally acquired TM of the MMF also obeyed the circular Gaussian distribution, it could be reasonable to use Eq.~(\ref{eq:ten} for normalizing the empirical focusing PBRs and calculating the focusing efficiency. The average focusing efficiencies (denoted by the crosses in the boxplots) were 16.45\%, 26.01\%, 37.46\%, and 55.17\% for prVBEM, GGS 2-1, RAF, and RAF 2-1 respectively. Surprisingly, RAF 2-1 achieved a considerably higher efficiency than RAF, \textit{i.e.}, $\sim17.72$\%, in comparison to the simulation results in Fig.~\ref{fig2}a. The reasons may include the noise type difference in the simulation and experiment, the larger TM dimension in the experiment (more challenging for recovery so that the gap is more obvious), etc. According to the boxplots of Fig 4a, quite a few foci approached or even surpassed the theoretical PBR for GGS 2-1, RAF, and especially RAF 2-1. However, their overall focusing efficiencies of 400 different foci on the working plane of MMF still saw a distance from the ideal level, using the retrieved TMs when $\gamma=5$.

\begin{figure}
  \centering
  \includegraphics[width=15cm]{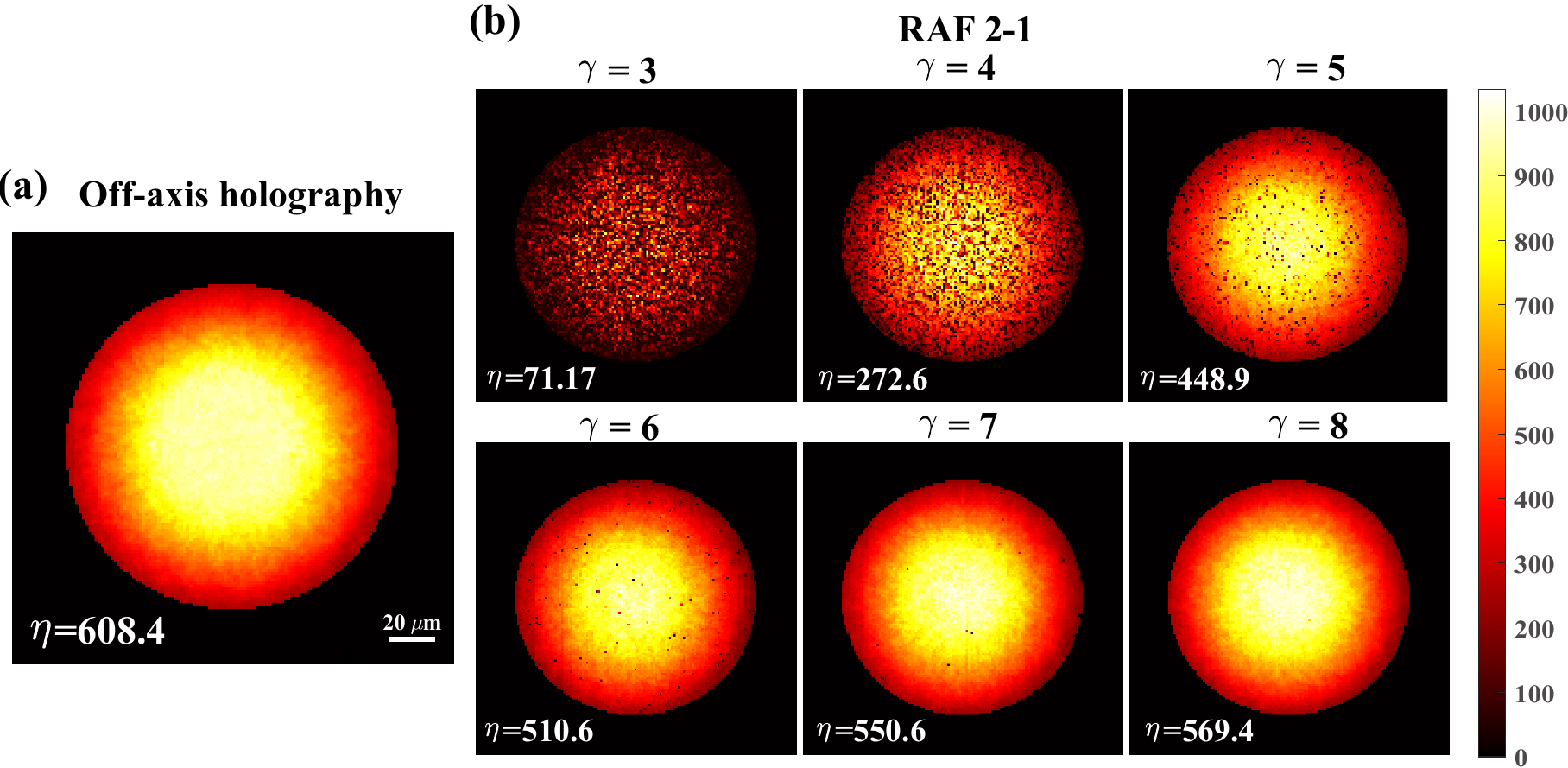}
  
  \caption 
  { \label{fig5}
  Comparison of the PBR maps of focusing on the working plane of the MMF using the TMs (a) measured by off-axis holography and (b) retrieved by RAF 2-1 under a range of $\gamma$ with $N=1024$. The scale bar in (a-b) was 20 $\mu m$.} 
\end{figure} 

In addition to single-spot focusing, multi-spot focusing experiment was also conducted under the same condition. This was achieved by superposing multiple phase-conjugate rows of the retrieved TM to construct a phase mask. The results of light focusing onto a pentagram pattern composed of 20 spots by different algorithms are shown in Fig.~\ref{fig4}b. The focusing uniformity were  = 33.7\%, -17.6\%, 40.0\%, and 69.2\%, respectively for the four algorithms. It can be observed that only RAF 2-1 produced a high-quality pentagram pattern by focusing light on all the pre-selected positions, thanks to its superior performance of TM retrieval. 

\begin{figure}
  \centering
  \includegraphics[width=15cm]{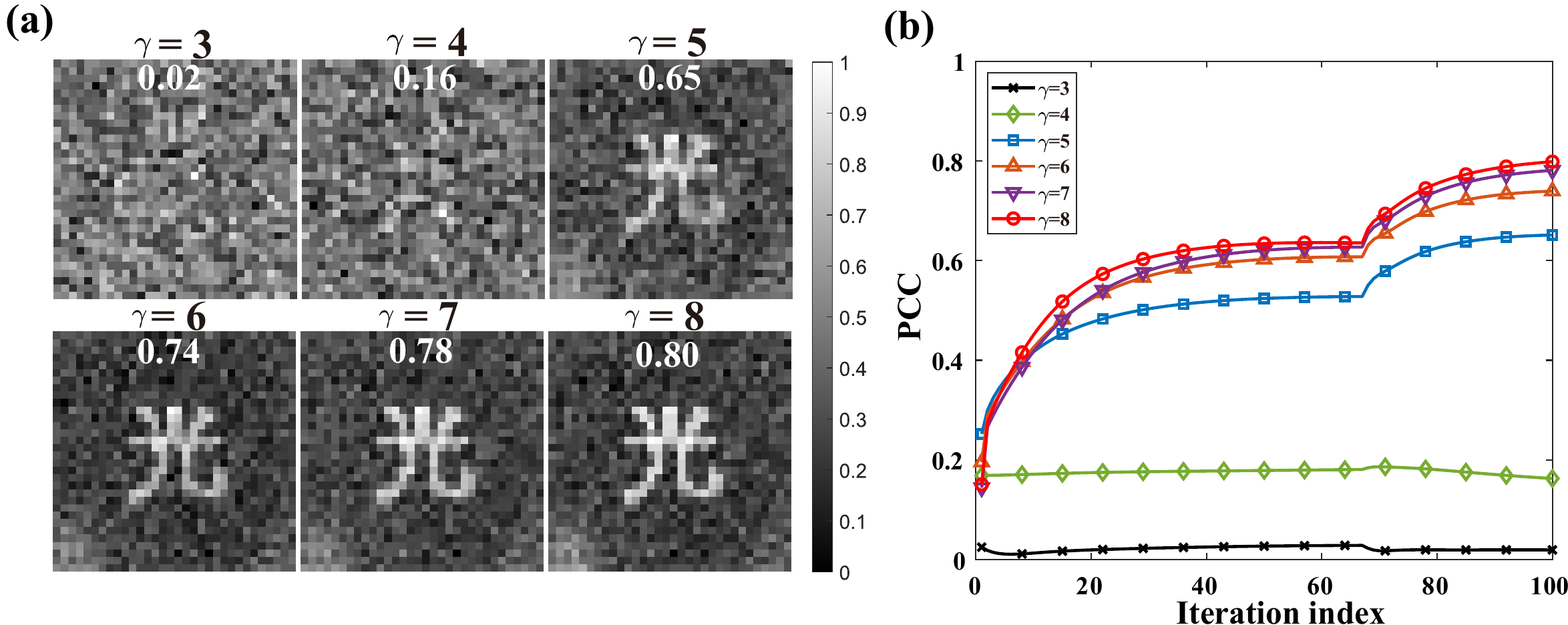}
  
  \caption 
  { \label{fig6}
  Comparison of image transmission results through MMF using the retrieved TMs by RAF 2-1 under a range of $\gamma$ with $N=1024$. (a) Normalized reconstructed images for an object of Chinese character, with the values of PCC to the object labeled (b) The progression curves of PCC during the iterative reconstruction.} 
\end{figure}

The accuracy of TM retrieval by our RAF 2-1 was further compared with the off-axis holography, \textit{i.e.}, the golden standard for the measurement of TM. To do so, we scanned across the whole fiber region on the working plane of the MMF, so that a map of focusing PBR could be synthesized, which was used to fully evaluate the accuracy of TM. The fiber region was determined by identifying the largest connected region in the binarized image taken when all pixels on the DMD were turned on. Using circular fitting of the fiber region, the center and radius of the fiber region were obtained, which were then used to create a binary mask of the fiber region. In the experiment, there were 8685 pixels inside the circular fiber region of the $140 \times 140$ output field, which correspond to 8685 rows of TM to be retrieved. Fig.~\ref{fig5} summarizes the results of focusing PBR maps with the TM measured by off-axis holography and the TMs retrieved by RAF 2-1 under a range of $\gamma$ from 3 to 8. The mean PBR by the golden standard method is 608.4. Compared with the theoretical PBR (\textit{i.e.}, $\eta=804$ ), it is reasonable given that the focusing quality degraded in the fiber fringe area due to the inhomogeneous mode excitation inside the MMF. As for RAF 2-1, there are many dark spots in the PBR map synthesized under small $\gamma$, indicating poor accuracy of the corresponding rows of TM being retrieved. With a larger $\gamma$, the PBR map becomes more homogeneous with less dark spots, which means an overall improvement of the TM accuracy. Notably, when $\gamma=8$, the PBR map by RAF 2-1 is comparable to that of the holography, with a mean PBR of 569.4 reaching $\sim93.6$\% efficiency of the golden standard experimentally. The key advantage is RAF 2-1 does not require holographic setup for the calibration of TM, compatible with more applications. Additionally, with parallel operation and GPU implementation, the TM retrieval process by RAF 2-1 was fast enough. For example, under $\gamma=8$, retrieving an 8685×1024 TM by RAF 2-1 averagely took 42.3 s, when using a computer with an Intel Xeon CPU E5-1650 v3 @3.50 GHz, a NVIDIA RTX2080 Ti GPU, and 128 GB RAM. 

Using the retrieved TM by RAF 2-1, one can further reconstruct an input object from intensity-only speckle measurement with one more phase retrieval. The reconstruction result relies heavily on the quality of the recovered TM, which acts as the measurement matrix. Fig.~\ref{fig6}a shows the results of reconstructing a $32 \times 32$ phase object of a Chinese character (meaning “light”), by taking 100 iterations with the TM of MMF retrieved by RAF 2-1 when $\gamma$ increased from 3 to 8. Pearson correlation coefficient (PCC) is used to quantify the similarity between the reconstructed image ${\boldsymbol{I}}_R$ and the ground truth ${\boldsymbol{I}}_G$, which is given by
\begin{equation}
\rm{PCC}=\frac{\left \langle \left (     \boldsymbol{I}_R-\left \langle \boldsymbol{I}_R  \right \rangle          \right ) \left (     \boldsymbol{I}_G-\left \langle \boldsymbol{I}_G  \right \rangle          \right ) \right \rangle}{\left \langle \left (     \boldsymbol{I}_R-\left \langle \boldsymbol{I}_R  \right \rangle          \right )^2 \right \rangle \left \langle \left (     \boldsymbol{I}_G-\left \langle \boldsymbol{I}_G  \right \rangle          \right )^2 \right \rangle}.
\label{eq:thirteen}
\end{equation}	 
The curves of PCC under different cases of $\gamma$ are also provided in Fig 6b. The upsurges of PCCs at 67th iteration confirm that the signal recovery is significantly boosted after the gradient heating in the first $2/3$ iterations. The final PCCs are: 0.02, 0.16, 0.65, 0.74, 0.78, 0.80 for $\gamma$ = 3, 4, 5, 6, 7, 8 respectively. As can be observed, the reconstructed image could be recognized starting from  $\gamma=5$ and attains the best quality when $\gamma=8$. To summarize, image transmission through the MMF is demonstrated with the proposed nonconvex approach, which further validates the accuracy of the retrieved TM.

\section{Discussion and conclusion}

There have been various phase retrieval algorithms used for solving the TM retrieval problem, as introduced earlier. RAF, as one of the best in the nonconvex family, has been reported previously \cite{ref53} to be highly competitive for image restoration from speckle measurement. To the best of our knowledge, we first adopted it for non-holographic calibration of TM \cite{ref41}. More importantly, our modified version, RAF 2-1, with an additional step of gradient heating, has shown remarkable improvement in the robustness against noise and the TM retrieval accuracy in both simulations and experiments. The numerical evaluation of RAF and RAF 2-1 can be further seen in Appendix B. Besides the above modification, we resort to speeding up the convergence of RAF for TM retrieval. Efforts include employing adaptive step size in the gradient descent process or other gradient acceleration schemes, such as Limited memory-BFGS (L-BFGS) \cite{ref54} and nonlinear conjugate gradient (NCG) \cite{ref55} methods. However, the improvements are not very impressive, with details referred to Appendix C.

There are also several limitations in the study. In the experimental setup, the MMF output field was relayed by a collimator instead of an objective lens. Consequently, the working plane of the MMF was immovable, which had a certain distance (about tens of micrometers) away from the fiber end. That said, the setup was sufficient for retrieving a reliable TM and focusing on the working plane for demonstration. An objective lens is needed only for measuring the TMs corresponding to different working planes. In addition, since there is phase ambiguity for the formulated LSE objective function in Eq.~(\ref{eq:four}), a phase offset exists for each row of the retrieved TM. However, it does not affect the intensity of the generated 2D foci. Further phase correction \cite{ref27} is indispensable when it comes to MMF 3D volumetric focusing and imaging.

In conclusion, we have proposed a modified nonconvex approach, RAF 2-1, for retrieving the TM of MMF based on speckle intensity measurements. Theoretically, RAF 2-1 can achieve optimum focusing efficiency with less running time or sampling rate than the previously reported TM retrieval methods. The experimental results of light control through a MMF confirm a comparable performance of RAF 2-1 to the golden standard holography method for TM measurement. RAF 2-1 is also computationally efficient that took averagely 42.3 s to recover an 8685×1024 TM ($\gamma=8$) on a regular computer under parallel operation and GPU implementation. Endowed with the advantages of optimum efficiency, fast execution, and a reference-less setup, RAF 2-1 allows for broad applications in MMF-based imaging, manipulation, and treatment etc.

\subsection* {Appendix A: The best iteration ratio for the two-step gradient iteration process of RAF 2-1}
\label{sect:appendix1}

\begin{figure}
    \centering
    \includegraphics[width=10cm]{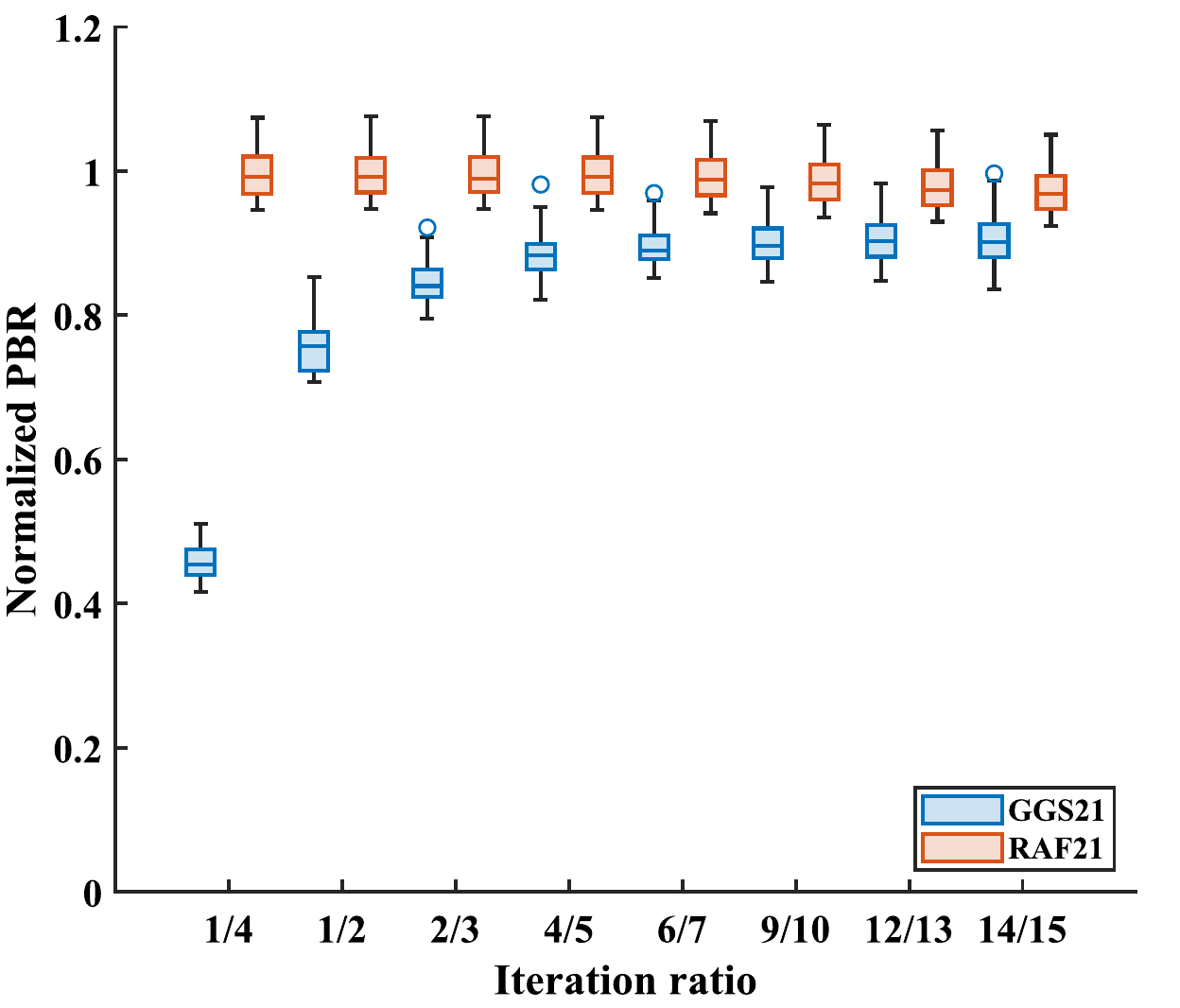}

    \caption 
    { \label{figS1}
    Normalized focusing PBR achieved by GGS 2-1 and RAF 2-1 under a series of iteration ratios during their two-step gradient iterations, respectively.} 
\end{figure}

There are two steps in the gradient iteration process regarding the proposed RAF 2-1. In order to determine the number of iterations in Steps 1 and 2 (with the total number fixed) for the best performance, numerical experiments were conducted to compare the focusing efficiency achieved by RAF 2-1 under a series of iteration ratios. Moreover, since GGS 2-1 also involved the two-step gradient descent, it inspires this work and used for performance comparison. Therefore, the best ratio of iteration of GGS should also be determined. Fig.~\ref{figS1} gives the results of both RAF 2-1 and GGS 2-1, with a total iteration of 300. It can be observed for RAF 2-1, there are minor differences of focusing efficiency among different iteration ratios, while the one at $2/3$ is chosen as the best iteration ratio due to a slight advantage. As for GGS 2-1, the best focusing efficiency is around an iteration ratio of $9/10$, which is also consistent with the original research that adopted 287 and 34 iterations in Steps 1 and 2 for GGS 2-1, respectively.

\subsection* {Appendix B: Numerical evaluation of RAF and RAF 2-1}
\label{sect:appendix2}

\begin{figure}
    \centering
    \includegraphics[width=10cm]{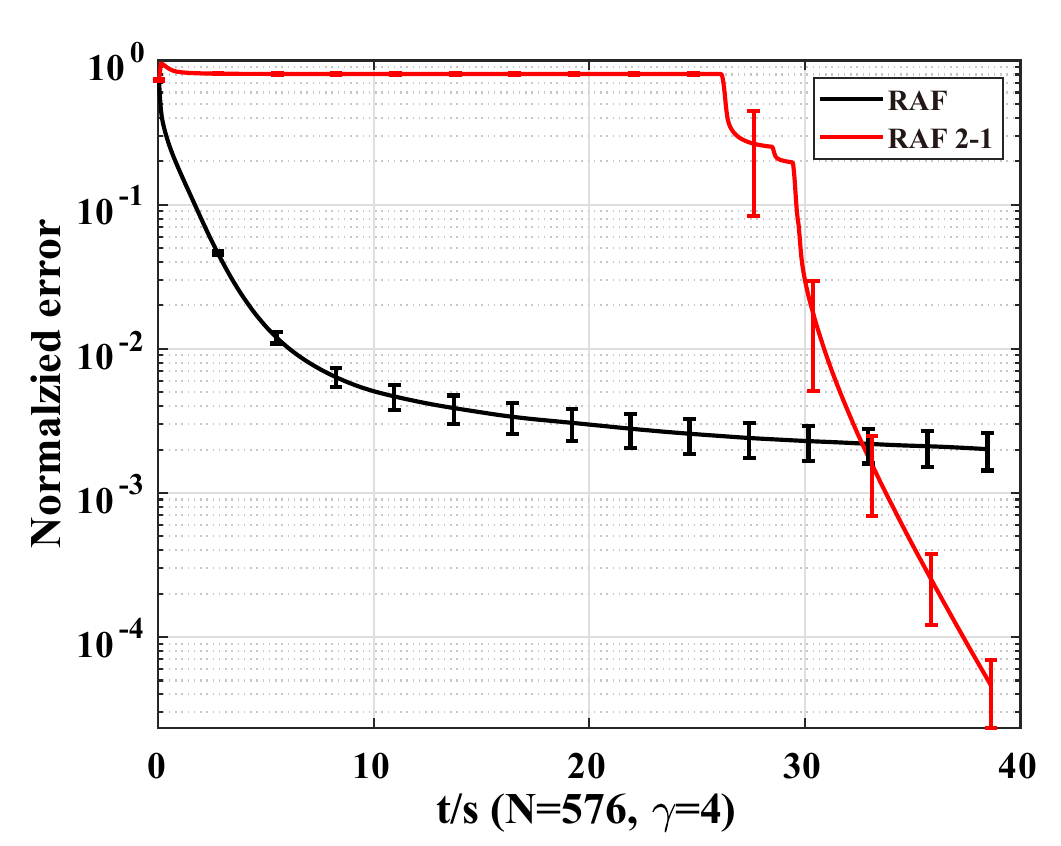}
    \caption 
    { \label{figS2}
    Normalized curves of error as a function of running time for RAF and RAF 2-1 when $N=576$ and $\gamma=4$. Note the error bars denote the standard deviations of 30 repeated tests.} 
\end{figure}

As mentioned in the Methods section, the modified version, RAF 2-1, is more effective for TM retrieval. To evaluate how the performance of RAF 2-1 is better than the original RAF, a numerical experiment in a noiseless condition was performed for retrieving the TM that corresponds to 400 output modes. The curves of the averaged errors after normalization are presented in Fig.~\ref{figS2}. The measurement error for the ${i^{th}}$ row of TM is defined as
\begin{equation}
  \mathop {\rm{error}_i} = \left\| {\left| {{{\boldsymbol{X}}^H}{{{\boldsymbol{\hat a}}}_i}} \right| - {{\boldsymbol{y}}_i}} \right\|_2^2, i = 1, \cdots, 400
  \label{eq:S1}
\end{equation}
where ${\boldsymbol{\hat a}}_i$ is the estimate of the ${i^{th}}$ row of TM and other notations are with the same meaning as in the Methods section. It can be observed that the error of RAF 2-1 can finally decline to a level of as low as $10^{-4}$, much lower than that of RAF. This indicates a more accurate result of TM retrieval by RAF 2-1. 

\subsection* {Appendix C: Accelerated gradient descent for RAF}
\label{sect:appendix3}

\begin{figure}
    \centering
    \includegraphics[width=14cm]{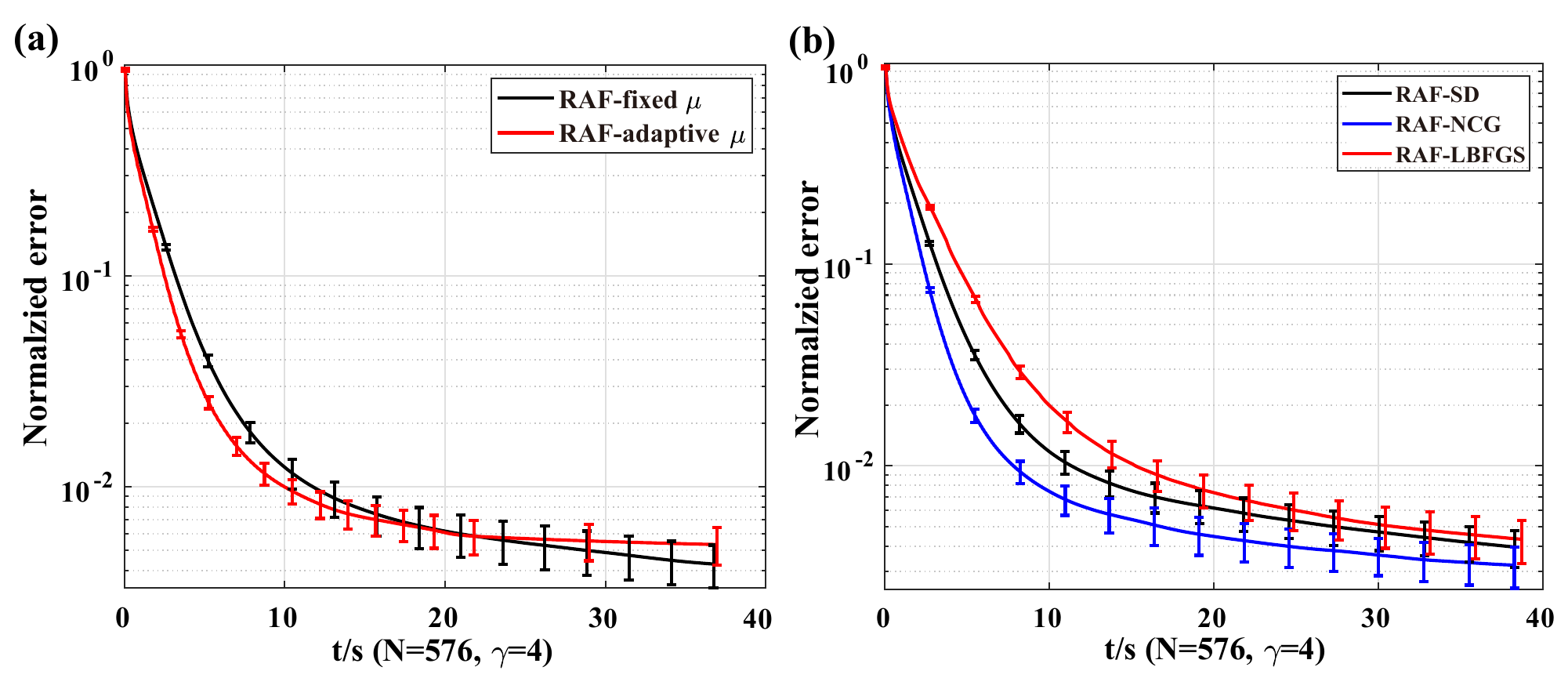}

    \caption 
    {\label{figS3}
    (a). Normalized curves of error as a function of running time for RAF with a fixed step size ($\mu$) or an adaptive one. (b) Normalized curves of error as a function of running time for RAF with steep descent (SD), non-linear conjugate gradient (NCG), or L-BFGS.} 
\end{figure}

As discussed earlier, the ways to accelerate the convergence of RAF were also studied, by using an adaptive step size and a more advanced gradient descent solver. First, we adopted the Barzilai-Borwein method for non-monotonic backtracking line-search of step size, which was compared with the fixed one ($\mu=3$). As seen in Fig.~\ref{figS3}a, the measurement error of using adaptive $\mu$ drops slightly more rapidly than that of fixed $\mu$ within the first 20 s of running time, while the latter eventually declines to a lower level. This shows the adaptive step size method is not very effective, although it could be better with parameter finetuning. As for the gradient descent solver, besides the regular steep descent using the negative first derivative (\textit{i.e.}, the gradient) as the descent direction, acceleration methods such as NCG and L-BFGS were employed for comparison. Since NCG and especially L-BFGS entail more computations per iteration than SD, for fair comparison, the curves of error as a function of running time (instead of iterations) for different optimization methods were compared, as shown in Fig.~\ref{figS3}b. We can see that NCG has the fastest convergence with the same running time. The reason that L-BFGS method is even slower in the declining trend of error is attributed to the far more seconds per iteration it requires. In fact, the average number of iterations taken by SD, NCG, and L-BFGS are 671, 656, and 411, respectively. The convergence for L-BFGS could be the fastest if compared from the perspective of error declining versus iterations. 

\subsection*{Disclosures}
The authors declare that there are no conflicts of interest fo this work.

\subsection* {Data Availability} 
The code demo of RAF 2-1 is available in \url{https://github.com/Ford666/RAF21_TM_retrieval}, and the data that support the findings of this study could be available by reasonable request to the corresponding author.

\subsection* {Acknowledgments}
This work is supported by the National Natural Science Foundation of China (NSFC) (81930048), Hong Kong Innovation and Technology Commission (GHP/043/19SZ, GHP/044/19GD), Hong Kong Research Grant Council (15217721, R5029-19, C7074-21GF), Guangdong Science and Technology Commission (2019BT02X105), Shenzhen Science and Technology Innovation Commission (JCYJ20220818100202005), and Hong Kong Polytechnic University (P0038180, P0039517, P0043485, P0045762). The authors would also like to thank the Photonics Research Institute of the Hong Kong Polytechnic University for facility support. \\

†These authors contributed equally.

\bibliographystyle{unsrtnat}
\bibliography{references}  

\end{document}